\documentclass[prl,twocolumn,showpacs]{revtex4}
\usepackage{amsmath}
\usepackage{graphicx}

\newcommand{\ket}[1]{\left|#1\right\rangle}
\begin{document}
\title{Adiabatic Quantum State Manipulation of
Single Trapped Atoms}
    \author{M.~Khudaverdyan}
    \author{W.~Alt}
    \author{I.~Dotsenko}
    \author{L.~F\"{o}rster}
    \author{S.~Kuhr}
    \author{D.~Meschede}
    \author{Y.~Miroshnychenko}
    \author{D.~Schrader}
    \author{A.~Rauschenbeutel}
    \email{rauschenbeutel@iap.uni-bonn.de}
    \affiliation{Institut f\"ur Angewandte Physik, Universit\"at
Bonn,
    Wegelerstr.~8, D-53115 Bonn, Germany}
\date{\today}

\begin{abstract}
We use microwave induced adiabatic passages for selective spin
flips within a string of optically trapped individual neutral Cs
atoms. We position-dependently shift the atomic transition
frequency with a magnetic field gradient. To flip the spin of a
selected atom, we optically measure its position and sweep the
microwave frequency across its respective resonance frequency. We
analyze the addressing resolution and the experimental robustness
of this scheme. Furthermore, we show that adiabatic spin flips can
also be induced with a fixed microwave frequency by
deterministically transporting the atoms across the position of
resonance.
\end{abstract}

\maketitle

Adiabatic passages (APs) \cite{1} are an interesting alternative
to purely resonant interaction for controlling the quantum state
of atoms. They rely on the fact that the coupled atom--field
system remains in its instantaneous eigenstate if the variation of
its parameters (atom--field detuning and field strength) is
sufficiently slow and smooth. One can thus adiabatically transfer
a system from an initial to a final state and, under certain
conditions, fluctuations of the parameters will not affect the
outcome of the AP. The pioneering works concerning APs were
performed in nuclear magnetism to achieve inversion of a spin
system \cite{2}. The first application of this method in the
optical domain was realized in \cite{3} to invert the population
in NH$_3$ molecules. Ever since, a multitude of different AP
techniques have been proposed and successfully realized \cite{4}.
It has also been shown that APs can be used to robustly prepare
superpositions of energy eigenstates \cite{5}. Furthermore, it was
proposed to prepare entanglement and implement quantum logic
operations through adiabatic processes \cite{5b,6}.

We have recently demonstrated that a string of neutral Cs atoms
stored in a standing wave optical dipole trap can be used to
realize a quantum register \cite{7}. There, quantum information
was written into the two Cs hyperfine ground states by subjecting
selected atoms to {\em resonant} microwave pulses. Here, we report
on the realization of an {\em adiabatic} method for flipping the
states of individual atoms out of a string. As in \cite{7}, the
atoms are discriminated through a position dependent transition
frequency. The APs are accomplished by sweeping the microwave
frequency across the resonance frequency of the respective atom.
We investigate the performance of our method by recording AP
spectra of few as well as of single atoms, yielding high-quality
data in perfect agreement with theory. The spatial discrimination
of this scheme is comparable to resonant addressing. At the same
time, the method is much more robust than in the resonant case. It
is therefore a useful tool for the manipulation and control of our
quantum register.

Combining {\em fixed frequency} microwave pulses with our
``optical conveyor belt'' technique \cite{8,9} inside a magnetic
field gradient, we furthermore realize APs by transporting atoms
across the position of resonance. In this experiment, the
atom--field coupling and the position dependence of the transition
frequency are chosen such that the dynamics of the system is
similar to that of atoms coupled to the mode of an optical high
finesse Fabry-Perot resonator \cite{10}. We show that our
transport procedure is sufficiently smooth to guarantee
adiabaticity over a large range of parameters, indicating that
motion-induced APs could effectively be used in our system to
control the atom--field interaction in cavity QED experiments.

\begin{figure}[ht]
          \centering
          \includegraphics [scale=0.36]{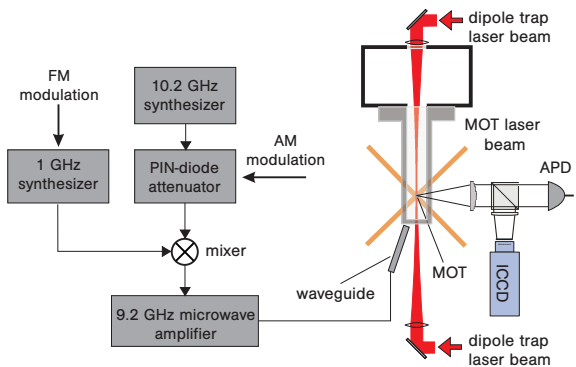}
          \caption{Experimental setup. See text for details.}
          \label{fig:Microwavesetup}
\end{figure}

We use a specially designed magneto-optical trap (MOT) as a source
of a well-defined number of cold cesium atoms. 
Fluorescence light from the MOT is collected by an objective lens
and is imaged onto the photocathode of an intensified CCD camera
(ICCD) and onto an avalanche photodiode (APD), see
Fig.~\ref{fig:Microwavesetup}. Whereas the APD signal allows us to
monitor the number of atoms, the ICCD provides us with information
about their positions \cite{12}. The atoms are transferred from
the MOT to a standing wave dipole trap \cite{8,9}, generated by
two counter-propagating far red-detuned laser beams with a
wavelength of $\lambda=1064$~nm. They interfere and produce a
chain of potential wells of typically 1~mK depth. Further details
can be found in previous publications \cite{9,13}.

As qubit states we employ the outermost Zeeman sublevels
$|F=4,m_F=4\rangle$ ($\ket{0}$) and $|F=3,m_F=3\rangle$
($\ket{1}$) of the two $6S_{1/2}$ hyperfine ground states with a
transition frequency $\omega_\mathrm{at}/2\pi$ near 9.2~GHz. In
order to spectroscopically distinguish atoms at different
positions along the dipole trap axis we apply a magnetic field
gradient of 13~G/cm along this axis. This results in a position
dependent transition frequency between states $\ket{0}$ and
$\ket{1}$ of $\partial_x \omega_\mathrm{at}/2\pi= 3.2$~kHz/$\mu$m
\cite{7}.

Each experimental sequence starts by loading atoms into the MOT
and transferring them to the dipole trap. In order to further
lower the temperature of the atoms in the trapping potential an
optical molasses is switched on shortly after the transfer. Then
the trap depth is lowered from 1~mK to 140~$\mu$K in order to
reduce the differential light shift (see below). A guiding
magnetic field of 4~G is applied along the dipole trap axis, which
shifts $\omega_\mathrm{at}/2\pi$ by 9.8~MHz. Optical pumping to
the state $\ket{0}$ is accomplished by exposing the atoms to a
$\sigma^+$-polarized laser beam along the guiding field, resonant
with the $F=4\leftrightarrow F'=4$ transition, together with a
repumping laser beam on the $F=3\leftrightarrow F'=4$ transition.

The microwave radiation for the APs is generated using two
frequency synthesizers (see Fig.~\ref{fig:Microwavesetup}). The
first one (Agilent 8375A) generates a signal at 10.2~GHz. Its
output is sent through a PIN-diode attenuator which allows a fast
modulation of the signal amplitude (AM). The second generator
(Rohde \& Schwarz SML02) is frequency modulated (FM) and operates
at 1~GHz. Both signals are mixed and the difference frequency at
9.2~GHz is amplified to 36~dBm. Finally, it is sent through an
open ended waveguide and directed to the trapped atoms. This
microwave setup combines the excellent frequency stability of the
10.2~GHz source with the phase continuous frequency sweeping
capacity of the 1~GHz source.

The computer generated FM and AM control signals allow us to
realize microwave pulses of arbitrary shapes. For inducing APs, we
have chosen the pulse shape suggested in \cite{14}:
\begin{equation}\label{e:adiabaticPulseShape}
 \begin{split}
\Omega_{\rm R}(t)&=\Omega_{\rm max}\sin^2(\pi t/t_{\rm
p}) \\
\delta(t)&=\delta_{\rm c} + {\rm sign}(t-t_{\rm p}/2)
\cdot\delta_{\rm max}\sqrt{1-\sin^4 (\pi t/t_{\rm p})}
 \end{split}
\end{equation}
for $0\leq t\leq t_{\rm p}$. Here, $\Omega_{\rm max}$ is the
maximum value for the Rabi frequency,
$\delta(t)=\omega(t)-\omega_\mathrm{at}$ is the detuning of the
microwave frequency from atomic resonance, $\pm\delta_{\rm max}$
is the span of the frequency sweep around the central detuning
$\delta_{\rm c}$, and $t_{\rm p}$ is the pulse duration.

The population transfer efficiency of the APs depends on the
adiabaticity of the frequency sweep, which is reflected by the
condition \cite{15}:
\begin{equation}\label{e:AdiabaticityCondition}
   \frac{|\dot{\delta}(t)\Omega_{\rm R}(t)-\delta(t)\dot{\Omega}_{\rm
    R}(t)|}{2(\Omega_{\rm R}(t)^2+\delta(t)^2)^{3/2}} \ll 1.
\end{equation}
The pulse shape of eq.~(\ref{e:adiabaticPulseShape}) fulfills this
adiabaticity condition over a wide range of the central detuning
$\delta_{\rm c}$.

\begin{figure}[t]
        \centering
                   {\includegraphics [scale=0.4]{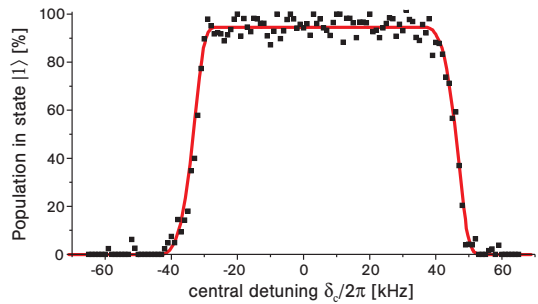}}
\caption{Spectrum of adiabatic population transfer. Each data
point shows the average population transfer to state $\ket{1}$ for
ten shots with about five atoms each. The solid line is a
theoretical fit, derived from optical Bloch equations. The maximum
population transfer efficiency of $95 \pm 0.4~\%$ is limited by
imperfect initialization of state $\ket{0}$, which depends
crucially on the polarization purity of the optical pumping laser.
The Bloch equations alone predict 100~$\%$ population transfer
efficiency. }
        \label{fig:AP_homogen}
\end{figure}

We first demonstrate adiabatic population transfer in a
homogeneous magnetic field. Atoms are loaded into the dipole trap
and are initialized in state $\ket{0}$. Then the frequency
modulated microwave pulse is applied. The final state of the atoms
is detected using a ``push-out'' laser beam \cite{16}. It expels
atoms in $\ket{0}$ out of the trap with 99~$\%$ efficiency, while
atoms in $\ket{1}$ remain trapped with probability higher than
99~$\%$. The final number of atoms is revealed by transferring
them back into the MOT and by detecting their fluorescence. The
recorded spectrum is presented in Fig.~\ref{fig:AP_homogen}, using
the following microwave pulse parameters: $\Omega_{\rm
max}/2\pi=28$~kHz, $\delta_{\rm max}/2\pi=40$~kHz, and $t_{\rm
p}=2$~ms. We have stepped the central detuning $\delta_{\rm
c}/2\pi$ from $-65$~kHz to 65~kHz in steps of 1~kHz.

The wide plateau in Fig.~\ref{fig:AP_homogen} shows a population
transfer efficiency $P_1>90~\%$ for $-30~{\rm kHz}<\delta_{\rm
c}/2\pi<40$~kHz. It is constant over a large interval of the
central detuning demonstrating robustness of the spin flip
efficiency with respect to frequency drifts and fluctuations.
Beyond this frequency interval the efficiency rapidly drops to
zero. The asymmetry of the spectrum is due to an inhomogeneous
broadening of the atomic resonance frequency which is caused by
the energy-dependent differential light shift $\hbar\delta_{\rm
ls}=\Delta E_0-\Delta E_1$ of the individual trapped atoms
\cite{16}, where $\Delta E_i$ is the light shift of level
$i=0$,~1, caused by the dipole trap laser.

We model the experimental data by calculating the transfer
efficiency $P_1(\delta_{\rm c})$ using Bloch equations for the set
of parameters used in the experiment. The thermal distribution of
atoms is then accounted for by convoluting $P_1(\delta_{\rm c})$
with the three-dimensional Boltzmann distribution of differential
light shifts \cite{16},
\begin{equation}\label{e:3dBoltzmannDistribution}
    p_\mathrm{B}(\delta_{\rm ls})=
    \frac{(\delta_{\rm ls}-\delta_{\rm ls}^{\rm max})^2}{2\delta_\mathrm{th}^3}
    \exp\left(-\frac{\delta_{\rm ls}-\delta_{\rm ls}^{\rm
    max}}{\delta_\mathrm{th}}\right)\ ,
\end{equation}
where $\delta_{\rm ls}^{\rm max}$ is the maximum differential
light shift and $\delta_\mathrm{th}$ is the change of the
differential light shift when the energy of the atom in the
trapping potential is increased by $k_B T$. The resulting function
\begin{equation}\label{e:adiabaticPoptransferWithLightShifts}
    \widetilde{P}_1(\delta_{\rm c})=
    P_\mathrm{max}\int_{\delta_{\rm ls}^{\rm max}}^{\,0}
    p_\mathrm{B}(\delta_{\rm ls})\,P_1(\delta_{\rm c}+\delta_{\rm
    ls})d\delta_{\rm ls}
\end{equation}
is then fitted to the data, with the maximum population
$P_\mathrm{max}=95$~\%, $\delta_\mathrm{th}/2\pi=1.7$~kHz, and
$\delta_{\rm ls}^{\rm max}/2\pi=-11$~kHz as fit parameters. The
fit (solid line in Fig.~\ref{fig:AP_homogen}) is in excellent
agreement with the experimental data.

In order to demonstrate position selective spin flips we now
switch on the magnetic field gradient. As in \cite{7} we load a
one atom on average into the dipole trap, determine its position
from a fluorescence image, and calculate the corresponding
resonance frequency $\omega_\mathrm{at}$. After initializing the
atom in state $\ket{0}$ we apply a microwave pulse of
shape~(\ref{e:adiabaticPulseShape}). We deliberately detune the
central microwave frequency from $\omega_\mathrm{at}$ by
$\delta_{\rm c}$. This corresponds to a position offset $\Delta
x=\delta_{\rm c}/\partial_x\omega_\mathrm{at}$. We detect the
hyperfine state after application of the microwave pulse and
record the single atom population transfer as a function of the
position offset $\Delta x$ along the trap axis.

\begin{figure}[t]
        \centering
            {\includegraphics [scale=0.4]{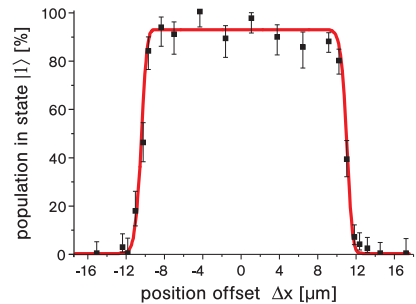}}
\caption{Position dependent adiabatic population transfer of
individual atoms in an inhomogeneous magnetic field. The graph
shows the population transfer induced by the AP as a function of
the position offset $\Delta x$ along the trap axis. Every data
point is obtained from about 40 single atom measurements. The
solid line is a theoretical fit of
eq.~(\ref{e:adiabaticPoptransferWithLightShifts}) with
$P_\mathrm{max}=93 \pm 1.7~\%$ again limited by the state
initialization procedure.}
        \label{fig:AP_gradient}
\end{figure}

The result of this measurement is presented in
Fig.~\ref{fig:AP_gradient}. Every data point is obtained from
about 40 post-selected single atom measurements. As expected, the
spectrum has a broad plateau with a population transfer efficiency
above 90~\%. At the edges this efficiency drops to zero within
$3~\mu$m. This addressing resolution is comparable to our resonant
addressing scheme presented in \cite{7}. However, the regime of
high-efficiency population transfer extends over an interval which
can be tuned by varying the span of the frequency sweep,
2$\delta_{\rm max}$, and which can thus be much larger ($18~\mu$m
in Fig.~\ref{fig:AP_gradient}) than in the resonant case,
providing robustness against frequency drifts and fluctuations.
Increasing the spatial resolution would require a larger pulse
duration $t_{\rm p}$ which however would decrease the transfer
efficiency due to dephasing mechanisms \cite{16}.

We now examine the possibility of inducing adiabatic spin flips
with a {\em fixed} microwave frequency by transporting the atoms
across the position of resonance with our optical conveyor belt
technique \cite{8,9}. This option is particularly interesting for
quantum information processing schemes in neutral atom cavity QED,
see e.~g.~\cite{17}. In order to scale up such schemes to a larger
number of atom-qubits, they will have to be shuttled into and out
of the interaction zone. If this transport is sufficiently smooth
to guarantee an adiabatic variation of the relevant parameters,
one could therefore envision to adiabatically modulate the
coupling strength of the atoms to the resonator by moving them
into and out of the mode. This technique could even be used to
reliably and robustly create entangled states through APs
\cite{5b,6}.

Here, we show that our transport realizes this adiabaticity
condition over a wide range of parameters. We move atoms in state
$\ket{0}$ along the dipole trap in the magnetic field gradient.
Before transporting the atoms, the microwave radiation is
adiabatically switched on. Then, during transportation, its
amplitude is kept constant, corresponding to a Rabi frequency of
$\Omega_{\rm R}/2\pi=26$~kHz. The initial atom positions are
distributed over $10~\mu$m, due to the size of the MOT \cite{12}.
Because of the magnetic field gradient this position spread
corresponds to a frequency spread of 32~kHz. We therefore tune the
microwave frequency 72~kHz to the red side of the
$\ket{0}\leftrightarrow\ket{1}$ transition, such that all atoms
are initially out of resonance. We then transport the atoms over a
distance of $d=132~\mu$m and thereby tune the atomic transition
frequency by $\partial_x\omega_\mathrm{at}\,d/2\pi=420$~kHz to the
red. These values are much larger than the initial detuning of the
atoms and the effective width $L=2\Omega_{\rm
R}/\partial_x\omega_\mathrm{at}=16~\mu$m of the region of
interaction. Therefore, after transport, all atoms, independently
of their initial positions, have crossed the region of resonance
and are again far detuned from the microwave frequency.

\begin{figure}[t]
        \centering
                    {\includegraphics [scale=0.26]{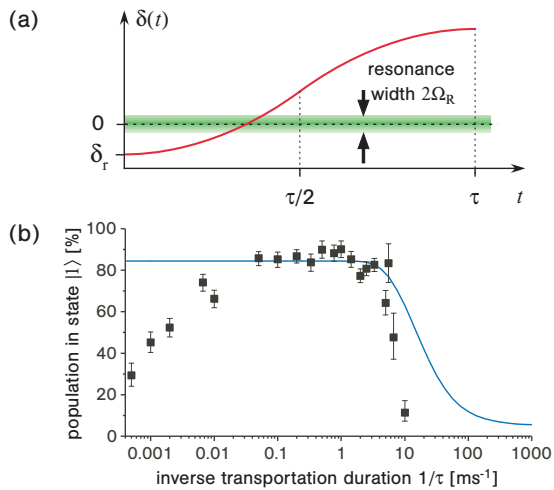}}
\caption{Transport induced adiabatic passages. (a) Sketch of the
temporal variation of the atom--field detuning during the
transport of duration $\tau$, see
eq.~(\ref{e:adiabaticPulseShapeTransport}). (b) Population in
state $\ket{1}$ after transport as function of $1/\tau$. Each data
point is an average over ten shots with about five atoms each. The
solid line is the theoretical population transfer, neglecting the
thermal motion of the atoms and decoherence of Rabi oscillations.
}
        \label{fig:AP_transport}
\end{figure}

The atoms are transported using a uniform acceleration $a$, which
undergoes a sign change at $d/2$. Neglecting the thermal motion of
the atoms inside the dipole trap, the resulting temporal variation
of the detuning is then (see Fig.~\ref{fig:AP_transport}~(a)):
\begin{equation}\label{e:adiabaticPulseShapeTransport}
\delta(t)= \left\{%
\begin{array}{ll}
  \delta_{\rm r} + a\,\partial_x\omega_\mathrm{at}\,t^{2}/2,
  & \mathrm{for} \ t\leq\tau /2 \\
  \delta_{\rm r} + a\,\partial_x\omega_\mathrm{at} \left(\frac{\tau^{2}}{4}-
  \frac{(t-\tau)^2}{2}\right), & \mathrm{for} \ t>\tau/2 \\
\end{array}%
\right. .
\end{equation}
Here, $\delta_{\rm r}$ is a random initial detuning due to the
spread of the atomic initial position, $a=4d/\tau^2$ is the
acceleration, $\partial_x\omega_\mathrm{at}$ is the gradient of
the atomic resonance frequency, and $\tau$ is the duration of
transportation.

In Fig.~\ref{fig:AP_transport}~(b) the percentage of atoms
transferred to state $\ket{1}$ after transport is shown as a
function of $1/\tau\propto v$, where $v$ is the average speed with
which the atoms cross the position of resonance. The population
transfer has been normalized to the transportation efficiency,
which has been measured independently, and exceeds 85~\% for
$\tau$ ranging from 300 $\mu$s to 20~ms. The maximum transfer
efficiency is most probably limited by imperfect state
initialization in this case. The solid line results from a
numerical simulation and is normalized to 85~\%. It predicts a
reduced transfer efficiency for $1/\tau > 3~\mathrm{ms}^{-1}$ due
to the loss of adiabaticity, in good agreement with the
experiment. For larger $1/\tau$-values, the experimental data
falls off faster than predicted by theory. This is most probably
due to the increasing excitation of axial oscillations of the
atoms inside the potential wells caused by the abrupt changes in
acceleration, which further reduce adiabaticity \cite{9}. For
$1/\tau < 0.05~\mathrm{ms}^{-1}$ the experimental data also falls
below the theoretical curve. This can be explained by the decay of
the driven atomic dipole caused by various dephasing mechanisms
\cite{16}.

Note that the effective width of the region of interaction $L$ is
comparable to the diameter of the Gaussian mode of typical
ultra-high $Q$ Fabry-Perot resonators for cavity QED experiments
\cite{10}. Furthermore, the most promising schemes for the
generation of entanglement in optical cavity QED rely on
four-photon Raman processes. Based on the proposal in \cite{17},
we have calculated an effective Rabi frequency of 18~kHz for this
Raman process using optimized experimental parameters. This value
is comparable with our microwave Rabi frequency $\Omega_{\rm R}$.
Our experimental results therefore indicate that our optical
conveyor belt allows us to transport two atoms sufficiently
smoothly through a resonator mode to realize entangling schemes
based on cavity-assisted APs.

Summarizing, we have demonstrated that microwave induced adiabatic
passages in a magnetic field gradient provide an efficient and
robust tool to manipulate the states of individual atoms in our
neutral atom quantum register. The spatial resolution of this
method is comparable with what can be achieved with resonant
pulses under similar experimental conditions \cite{7}, while
offering reduced sensitivity to fluctuations of experimental
parameters. We have furthermore shown that adiabatic population
transfer can also be realized with a fixed microwave frequency and
amplitude by transporting the atoms across the resonance position
using our optical conveyor belt technique. The wide range of
transportation times for which efficient transfer occurs
demonstrates that this scheme could indeed be used to
adiabatically modulate the coupling of the atoms to spatially
varying external fields. In particular in the context of cavity
quantum electrodynamics, adiabatic entangling schemes, induced by
the deterministic transport of two atoms through the resonator
mode, appear to be within the scope of our experiment.

This work was supported by the Deutsche
For\-schungs\-ge\-mein\-schaft (SPP 1078) and the EC (IST / FET /
QIPC project ``QGATES''). I.~D.~acknowledges funding from INTAS.
D.~S.~acknowledges funding by the Deutsche Telekom Stiftung.

\end{document}